\documentclass[journal=aesccq,manuscript=review]{achemso}

\usepackage{chemformula} 
\usepackage[T1]{fontenc} 

\author{Ana L\'opez-Sepulcre}
\affiliation{Univ. Grenoble Alpes, IPAG, F-38000 Grenoble, France}
\alsoaffiliation{Institut de Radioastronomie Millim\'etrique (IRAM), 300 rue de la piscine, F-38406 Saint-Martin d'H\`eres, France}
\email{lopez@iram.fr}
\author{Nadia Balucani}
\affiliation{Dipartimento di Chimica, Biologia e Biotecnologie, Universit\`a degli Studi di Perugia, I-06123 Perugia, Italy}
\author{Cecilia Ceccarelli}
\affiliation{Univ. Grenoble Alpes, IPAG, F-38000 Grenoble, France}
\author{Claudio Codella}
\affiliation{INAF, Osservatorio Astrofisico di Arcetri, Largo E. Fermi 5, I-50125, Firenze, Italy}
\alsoaffiliation{Univ. Grenoble Alpes, IPAG, F-38000 Grenoble, France}
\author{Fran\c cois Dulieu}
\affiliation{Universit\'e de Cergy-Pontoise, Observatoire de Paris, PSL University, Sorbonne Universit\'e, CNRS, LERMA, F-95000, Cergy-Pontoise, France}
\author{Patrice Theul\'e}
\affiliation{Aix-Marseille Universit\'e, PIIM UMR-CNRS 7345, 13397 Marseille, France}

\title[Interstellar formamide (NH$_2$CHO)]
  {Interstellar formamide (NH$_2$CHO), a key prebiotic precursor}

\abbreviations{}

\begin{document}

\begin{tocentry}
%
%

\centering
    \includegraphics[width=5.5cm]{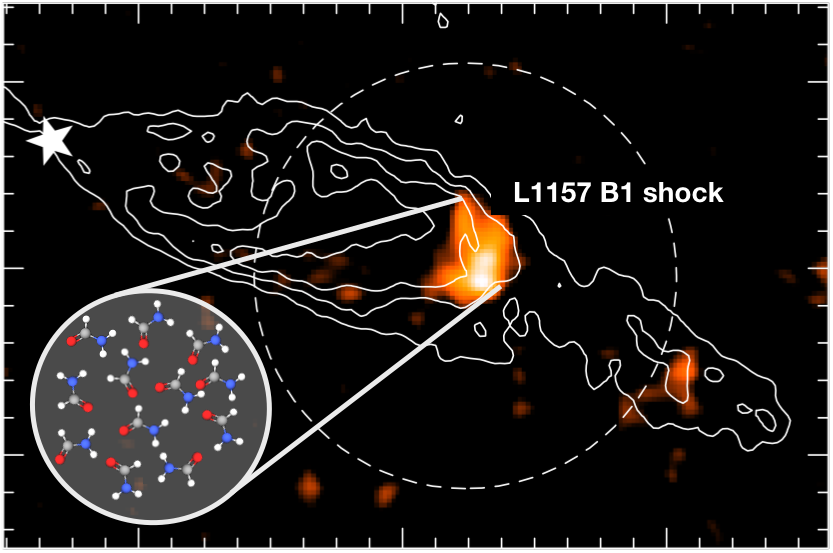}

%
%
\end{tocentry}

\begin{abstract}
Formamide (NH$_2$CHO) has been identified as a potential precursor of a wide variety of organic compounds essential to life, and many biochemical studies propose it likely played a crucial role in the context of the origin of life on our planet. The detection of formamide in comets, which are believed to have --at least partially-- inherited their current chemical composition during the birth of the Solar System, raises the question whether a non-negligible amount of formamide may have been exogenously delivered onto a very young Earth about four billion years ago. A crucial part of the effort to answer this question involves searching for formamide in regions where stars and planets are forming today in our Galaxy, as this can shed light on its formation, survival, and chemical re-processing along the different evolutionary phases leading to a star and planetary system like our own.

The present review primarily addresses the chemistry of formamide in the interstellar medium, from the point of view of (i) astronomical observations, (ii) experiments, and (iii) theoretical calculations. While focusing on just one molecule, this review also more generally reflects the importance of joining efforts across multiple scientific disciplines in order to make progress in the highly interdisciplinary science of astrochemistry.
\end{abstract}

\section{Introduction: why do we care about formamide?}

The mysterious nature of life has fascinated humankind since ancient times. How and when did life originate? Where is the limit between living and inert? Why does life thrive so easily on Earth while it appears to be a challenge elsewhere? Questions like these are still far from being answered today, despite them being the subject of numerous studies spanning a large range of scientific and philosophical disciplines.

While life may be regarded as something extremely complex, there is a surprising simplicity to it. Indeed, every single living being is composed of the same basic building blocks: organic molecules, whose elemental budget contains mostly carbon (C), hydrogen (H), oxygen (O), and nitrogen (N), with tiny quantities of other heavy elements such as sulphur (S) and phosphorus (P). Not only are all these elements easily found anywhere in the Universe, but they can also be found in the form of organic molecules (e.g. CH$_3$OH, HCOOCH$_3$, CH$_3$NH$_2$, CH$_3$CH$_2$CN) in certain space environments. This naturally leads to the following question: how does organic chemistry work in space? Answering this is one of the scopes of astrochemistry, and it may help us understand to what extent organic matter may have been exogenously delivered onto the young Earth.

Of all the organic molecules discovered in space so far, formamide (NH$_2$CHO) appears to be particularly promising when it comes to addressing the question of the emergence of life on Earth, for several reasons. First, it has an amide functional group (--N--C($=$O)--), which is necessary to form chains of amino acids and build up proteins. Second, it has been identified as a key precursor of a large variety of prebiotic molecules\cite{saladino12, saitta14}. Indeed, in the presence of an energy source, it promotes the synthesis of adenine, guanine, cytosine, and uracil, which are the four nucleobases of ribonucleic acid or RNA\cite{ferus15}, but it is also a precursor of carboxylic acids, amino acids, and sugars\cite{botta18}. Such a source of energy may have been provided about 4 -- 4.5\,billion years ago during planet formation, when the flux of asteroids and comets onto a very young Earth was orders of magnitude larger than it is today. Finally, formamide is liquid under typical terrestrial surface temperature and pressure conditions. This makes it an alternative solvent to water (H$_2$O) in the early days of prebiotic chemistry, as water can rapidly hydrolyse nucleic acids and proteins\cite{saladino12, adam18}.

In summary, the chemical versatility of formamide can lead, with the right catalysts, to the synthesis of many molecules that are key constituents of living organisms. In other words, it may provide a simple and unitary answer to the long-standing question ``Metabolism first or genetics first?''\cite{saladino12}.

The numerous chemical routes involved in the synthesis of biomolecules from formamide belong to the area of biochemistry. The present review tackles the opposite side of the story, namely the chemical pathways \emph{leading to} formamide, not on Earth, but in an environment which is much less chemically inviting: the interstellar medium (ISM). This is addressed from the point of view of observations (Sect.\,\ref{obs}), experiments (Sects.\,\ref{solid} and \ref{gas}), and theory (Sect.\,\ref{theory}). A summary and some concluding remarks are presented in Sect.\,\ref{final}.

\section{Observations of formamide in space}\label{obs}

\subsection{The early days}

The history of formamide observations in space dates back to almost 50 years ago. Indeed, its first detection in space was reported in 1971 by Rubin et al.\cite{rubin71}, who pointed the 140-foot telescope of the National Radio Astronomy Observatory (NRAO) towards the Sagittarius\,B2 region (hereafter Sgr\,B2), near the Galactic centre, and clearly observed the three $\Delta F = 0$ hyperfine components of the 2$_{1,1}-2_{1,2}$ rotational transition of NH$_2$CHO at 4.62\,GHz (6.5\,cm). Sgr\,B2 is the most massive star formation region (SFR) in our Galaxy. It exhibits an extraordinary molecular richness\cite{belloche13,alvaro17} and is a frequent target to search for new molecular species in the Galaxy\cite{belloche13,belloche14}. It therefore comes as no surprise that formamide was first detected in this region. This actually marked the first detection of an interstellar molecule containing C, H, O, and N.

Further observations with the same telescope and with the 64 m telescope at Parkes, coupled with spectroscopic work in the laboratory, confirmed the detection of formamide at 6.5 and 19\,cm wavelengths\cite{palmer71,ribes73,got73}. In particular, Gottlieb and collaborators obtained in 1973 the first formamide map in Sgr\,B2, with four pointings, allowing them to constrain the emission area to a diameter $<10'$ around the centre of the region \cite{got73}.

Detections of the $^{13}$C isotopologue of formamide were also reported in the literature early on\cite{laz78,gardner80,halfen17}. On the other hand, the detection of the $^{15}$N and deuterated isotopologues had to wait for the advent of the very sensitive (sub-)mm Atacama Large Millimeter and sub-millimeter Array (ALMA)\cite{belloche17,coutens16}.

Numerous observational works followed after the first detections that confirmed the presence of NH$_2$CHO in Sgr\,B2, at different mm and cm wavelengths and with a variety of telescopes, including single dishes and, more recently, interferometers\cite{hollis83,cum86,num98,num00,jones13,neill14}.

Almost 20\,years passed from the first formamide detection before it was observed elsewhere, far from the extreme environment around the Galactic centre. A 3\,mm single-dish spectral line survey performed by Turner\cite{turner89,turner91,sutton95} in the Orion Molecular Cloud 1 (OMC-1) revealed that NH$_2$CHO exists in this region. At a distance of about 390\,pc\cite{kounkel17}, the Orion molecular complex hosting OMC-1 and the iconic Orion Nebula, is the nearest high-mass SFR to the Sun. Since then, many other \textit{high-mass molecular cores} have been found to contain NH$_2$CHO in the gas phase\cite{gibb00,bisschop07,isokoski13,crockett15,suzuki18}, most of them as part of multi-line spectral surveys (see Table\,\ref{tsou}). However, high-mass molecular cores are subject to larger temperatures, densities, and ionising/dissociating irradiation than other, more quiet and lower-mass SFRs. In addition, their large distances and clumpy nature results in considerable confusion in the observations. What can we say about formamide in a region where a solar-mass star is forming instead? This is, after all, what we are most interested in if we want to make connections with the origin of our own Solar System.

\subsection{Formamide across solar-mass star formation}

The presence of formamide in cometary gas and ices constitutes a first hint that formamide likely existed in relatively large amounts in the Proto-Solar Nebula, prior to the formation of Solar System planets --including the Earth. Hale-Bopp was the first comet where NH$_2$CHO was detected, in its coma, via mm observations with the Caltech Submillimeter Observatory (CSO)\cite{lis97}. A non-negligible abundance of $(1-8) \times 10^{-4}$ with respect to H$_2$O was estimated. More recently, two more comets were found to contain gaseous formamide with similar amounts: C/2012 F6 (Lemmon) and C/2013 R1 (Lovejoy)\cite{biver14}. Finally, as part of the Rosetta mission, the Cometary Sampling and Composition (COSAC) mass spectrometer detected formamide in situ on comet 67P/Churyumov-Gerasimenko, together with other organic compounds\cite{goesmann15}.

Comets are believed to contain the most pristine material in the Solar System, and many efforts have been made so far to determine to what extent comets have inherited their contents (ices, silicates, etc.) from the early stages of Solar System formation. A key part of these efforts consists of comparing the ice/gas comet composition to that of current star birthplaces in our Galaxy (see e.g. the review by Caselli \& Ceccarelli\cite{cc12}).

In this context, it is natural to wonder whether formamide can be easily observed in solar-type birthplaces, and if so, at what evolutionary stage and under what physical conditions. Before answering this, we briefly recall the different evolutionary phases involved in \textit{low-mass star formation}, i.e. that of a star and planetary system like our own. The earliest phase is represented by a \textit{pre-stellar core}, which is a cold ($T < 10$\,K) and moderately dense ($n \sim 10^5$\,cm$^{-3}$) condensation of molecular gas and dust where matter slowly accumulates towards its centre. Condensation of CO and other molecules onto dust grains and hydrogenation of atoms and molecules on grain surfaces take place, together with an increase of molecular deuteration. The \textit{protostellar} phase follows, in which the gravitational energy due to a faster collapse towards the centre causes the inner core envelope to warm up, leading to sublimation of the icy mantles recovering dust grains. Disk-like accreting structures begin to form, and violent ejections of material in the form of bipolar jets occur in this phase. In the subsequent \textit{protoplanetary disk} phase, the envelope surrounding the protostellar object dissipates, and only the disk remains, whose dust grains will slowly coagulate to form planetesimals, and eventually, planets and other small bodies around the newly born star.

Today, we know from direct observational detections that formamide can be present in low-mass molecular cores at least as early as the protostellar stage. Kahane et al. \cite{kahane13} reported in 2013 the first detection of formamide in the solar-mass protostellar core IRAS\,16293--2422 in 2013, using the IRAM\,30-m telescope as part of TIMASSS (The IRAS 16293--2422 Millimeter And Submillimeter Spectral Survey)\cite{caux11}. IRAS\,16293--2422 is a well-studied binary system, known for its richness in saturated organic molecules such as methyl formate (HCOOCH$_3$), dimethyl ether (CH$_3$OCH$_3$), and ethanol (C$_2$H$_5$OH). These molecules are typically termed interstellar complex organic molecules (iCOMs). They exist in the inner compact ($< 100$\,au) and hot ($> 100$\,K) regions of some protostellar cores, known as \textit{hot corinos}\cite{cc99,cc07}, where the higher temperatures cause the sublimation of icy mantles onto the gas phase. This causes the release of many molecules previously trapped in the ices and triggers gas-phase chemical reactions that further enrich the molecular gas contents. It is therefore not surprising that formamide has been detected in IRAS\,16293--2422 alongside numerous other iCOMs, as it is a prototypical hot corino source.

Since then, the quest for formamide in other low-mass sources has not ceased. As can be seen in Table\,\ref{tsou}, several other low-mass molecular cores have been found to contain formamide in the gas phase using both single-dish and interferometric facilities\cite{yamaguchi12,mendoza14,ls15,taquet15,imai16,oya17,marcelino18}, even though the number is still quite small ($<\,10$ objects). From these few discoveries, there are several conclusions that can be extracted, which can put constraints on the formation of NH$_2$CHO in solar-mass star formation environments:

\begin{enumerate}
\item \textbf{Formamide abounds in hot corinos:} Formamide is detected mostly in protostellar cores hosting hot corinos, both young\cite{marcelino18} ($\sim 10^{4}$\,yr) and evolved\cite{ls15,bianchi19} ($\sim 10^{5}$\,yr), and including those of intermediate-mass, such as Cep\,E\cite{ls15,juan18}. In other words, the molecular gas containing formamide is hot ($T\,>\,100$\,K) and enriched in other iCOMs as a result of the sublimation of icy dust mantles.
\item \textbf{Large amounts of formamide in protostellar shocks:} Aside from hot corinos, the only other kind of low-mass star-formation environment where formamide has been discovered is in shock regions caused by protostellar jets, where dust grains are shattered and a rich gas chemistry is present. Only the outflow-shocked regions associated to two protostellar sources, L1157 and IRAS\,4A, have so far been reported to contain formamide\cite{yamaguchi12,mendoza14,codella17,ceccarelli17} (see Sect.\,\ref{shocks}).
\item \textbf{Not all protostars contain formamide:} The fact that formamide has been observed in a number of hot corinos does not imply that it is present in all protostars. The reason is simple: not all protostars contain a hot corino\cite{sakai13}. In fact, a few protostars without hot corinos have been searched for formamide without success. A noteworthy case is that of L1157, whose outflow shocks are rich in formamide, as mentioned above, but the protostar itself is devoid of emission from this molecule, as well as from other iCOMs typical of hot corinos\cite{ls15,lefloch18}.
\item \textbf{Formamide remains undetected in pre-stellar cores:} While more dedicated studies are needed, formamide does not appear to be present in detectable levels in pre-stellar cores, i.e. the phase preceding the protostellar one, or more generally, in cold molecular gas ($T < 10$\,K) even in sources where other iCOMs have been detected\cite{bacmann12,vastel14,js16}. Nevertheless, there is indirect evidence --through radiative transfer modelling of single-pointing single-dish observations-- that it may exist in the cold ($< 30$\,K) external envelope surrounding the hot corino binary IRAS 16293--2422\cite{jaber14} and Cep\,E\cite{ls15}, although a direct detection, i.e. with mapping techniques, would be needed to confirm this.
\item \textbf{No formamide is detected in protoplanetary disks:} This is to be expected, as very few organic species have been observed in this kind of environments\cite{walsh14,oberg15,favre18}. This is due to their compact sizes, cold mid-plane temperatures, and the absence of a substantial gas envelope, which altogether demand extremely high sensitivity observations in order to study their molecular gas composition. A recent modelling study predicts that fomamide mostly remains frozen onto dust grains in these regions and requires high temperature ($> 200$\,K) zones to be present in the gas\cite{quenard18b}.
\end{enumerate}

In short, formamide appears to be most abundant in regions where dust grains are disrupted by shocks or their icy mantles are released into the gas phase due to high temperatures. This does not necessarily imply that it is synthesised in the solid phase and then sublimated, as will be discussed below (Sect.\,\ref{shocks} and \ref{theory}).

\begin{table}
  \caption{Astronomical sources where formamide (NH$_2$CHO) has been detected}
  \label{tsou}
  \begin{tabular}{llll}
\hline
Source name  & Type & Telescope\textsuperscript{\emph{a}} & Reference(s)\textsuperscript{\emph{a}} \\
\hline
\hline
\multicolumn{4}{c}{Star formation regions}\\
\hline
Sgr\,B2 & high-mass & 140-foot NRAO & Rubin et al. \cite{rubin71}\\
Sgr\,A & high-mass & 140-foot NRAO & Rubin et al. \cite{rubin71}\\
OMC-1/Ori\,KL & high-mass & 11-m NRAO & Turner\cite{turner89}\\
G327.3--0.6 & high-mass & SEST & Gibb et al. \cite{gibb00}\\
G24.78 & high-mass & JCMT & Bisschop et al. \cite{bisschop07}\\
G75.78 & high-mass & JCMT & Bisschop et al. \cite{bisschop07}\\
NGC 6334\,IRS1 & high-mass & JCMT & Bisschop et al. \cite{bisschop07}\\
NGC 7538\,IRS1 & high-mass & JCMT & Bisschop et al. \cite{bisschop07}\\
NGC 7538\,S & high-mass & IRAM\,PdBI & Feng et al. \cite{feng16}\\
W\,3(H$_2$O) & high-mass & JCMT & Bisschop et al. \cite{bisschop07}\\
W\,33A & high-mass & JCMT & Bisschop et al. \cite{bisschop07}\\
IRAS 20126+4104\textsuperscript{\emph{b}} & high-mass & JCMT & Isokoski et al. \cite{isokoski13}\\
IRAS 18089--1732 & high-mass & JCMT & Isokoski et al. \cite{isokoski13}\\
G31.41+0.31 & high-mass & JCMT & Isokoski et al. \cite{isokoski13}\\
G10.47+0.03 & high-mass & NRO\,45-m & Suzuki et al. \cite{suzuki18}\\
G19.61--0.23 & high-mass & NRO\,45-m & Suzuki et al. \cite{suzuki18}\\
G34.3+0.2 & high-mass & NRO\,45-m & Suzuki et al. \cite{suzuki18}\\
W51\,e1/e2 & high-mass & NRO\,45-m & Suzuki et al. \cite{suzuki18}\\
G20.08--0.14N & high-mass & SMA & Xu et al. \cite{xu13}\\
G35.20 & high-mass & ALMA & Allen et al. \cite{allen17}\\
G35.03 & high-mass & ALMA & Allen et al. \cite{allen17}\\
G34.43+00.24\,MM3 & high-mass & ALMA & Sakai et al. \cite{sakai18}\\
Cep\,E & IM hot corino & IRAM\,30-m & L\'opez-Sepulcre et al. \cite{ls15}\\
OMC-2\,FIR\,4 & protocluster & IRAM\,30-m & L\'opez-Sepulcre et al. \cite{ls15}\\
Barnard\,1\textsuperscript{\emph{c}} & young hot corino & ALMA & Marcelino et al. \cite{marcelino18}\\
IRAS 16293--2422 & hot corino & IRAM\,30-m & Kahane et al. \cite{kahane13}\\
NGC\,1333\,IRAS\,4A & hot corino & IRAM\,30-m & L\'opez-Sepulcre et al. \cite{ls15}\\
NGC\,1333\,IRAS\,2A & hot corino & IRAM\,PdBI & Taquet et al. \cite{taquet15}\\
B335 & hot corino & ALMA & Imai et al. \cite{imai16}\\
L483 & hot corino & ALMA & Oya et al. \cite{oya17}\\
HH\,212 & hot corino & ALMA & Lee et al. \cite{lee17}\\
NGC\,1333\,SVS13A & evolved hot corino & IRAM\,30-m & L\'opez-Sepulcre et al. \cite{ls15}\\
L1157-B1 & protostellar shock & NRO\,45-m & Yamaguchi et al. \cite{yamaguchi12}\\
L1157-B2 & protostellar shock & IRAM\,30-m & Mendoza et al. \cite{mendoza14}\\
IRAS\,4\,As & protostellar shock & NOEMA & Ceccarelli et al. \cite{ceccarelli17}\\
\hline
\multicolumn{4}{c}{Solar System}\\
\hline
Hale-Bopp & comet & CSO & Bockel\'ee-Morvan et al. \cite{bm97}\\
C/2012 F6 (Lemmon) & comet & IRAM\,30-m & Biver et al. \cite{biver14}\\
C/2013 R1 (Lovejoy) & comet & IRAM\,30-m & Biver et al. \cite{biver14}\\
67P/Churyumov-Gerasimenko & comet & COSAC (Rosetta) & Goesmann et al. \cite{goesmann15}\\
\hline
\multicolumn{4}{c}{Other environments}\\
\hline
PKS 1830--211 & spiral galaxy & ATCA & M\"uller et al. \cite{muller13}\\
l.o.s. Sgr\,B2\textsuperscript{\emph{d}} & translucent cloud & ATCA & Corby et al. \cite{corby15}\\
CK Vulpeculae\textsuperscript{\emph{e}} & BD-WD merger & ALMA & Eyres et al. \cite{eyres18}\\
\hline
  \end{tabular}
\textsuperscript{\emph{a} Reference for the first detection only;} \textsuperscript{\emph{b} Tentative detection of a single line;} \textsuperscript{\emph{c} Tentative detection, with only one unblended line;} \textsuperscript{\emph{d} Detected in absorption along the line of sight towards Sgr\,B2;}  \textsuperscript{\emph{e} Detected in the disk surrounding a brown dwarf-white dwarf merger.} 
\end{table}

\subsection{How is formamide synthesised? The observational view}\label{interpretation}

From the beginning of the present decade, formamide has received considerable attention, surely triggered by the numerous biochemical studies that discuss its potential as a precursor of biological compounds essential to life. As a result, in the past few years, several publications have been specifically focused on astronomical observations of this molecular species, and on what such observations tell us about its formation in the ISM\cite{halfen11,kahane13,mendoza14,ls15,coutens16,codella17}. We discuss below three different directions that have so far been considered by observational astronomers with the aim of gaining more understanding on the formation of NH$_2$CHO: (i) abundance correlations, (ii) molecular deuteration, and (iii) protostellar shocks as laboratories to test the feasibility of certain chemical reactions.

\subsubsection{Abundance correlations}\label{corr}

\begin{figure} 
    \centering
    \begin{tabular}{lr}
    \includegraphics[width=8cm]{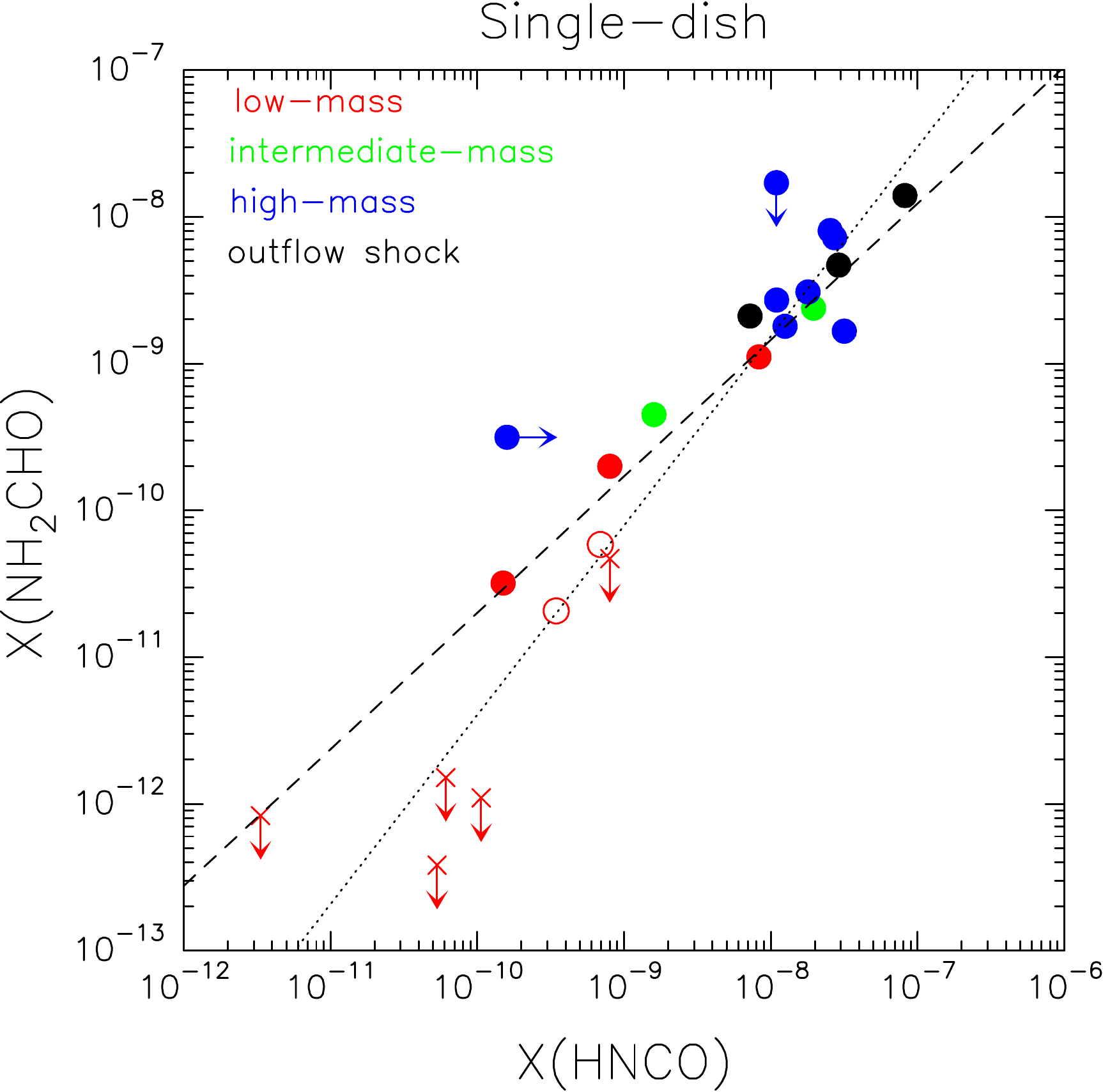} & \includegraphics[width=7cm]{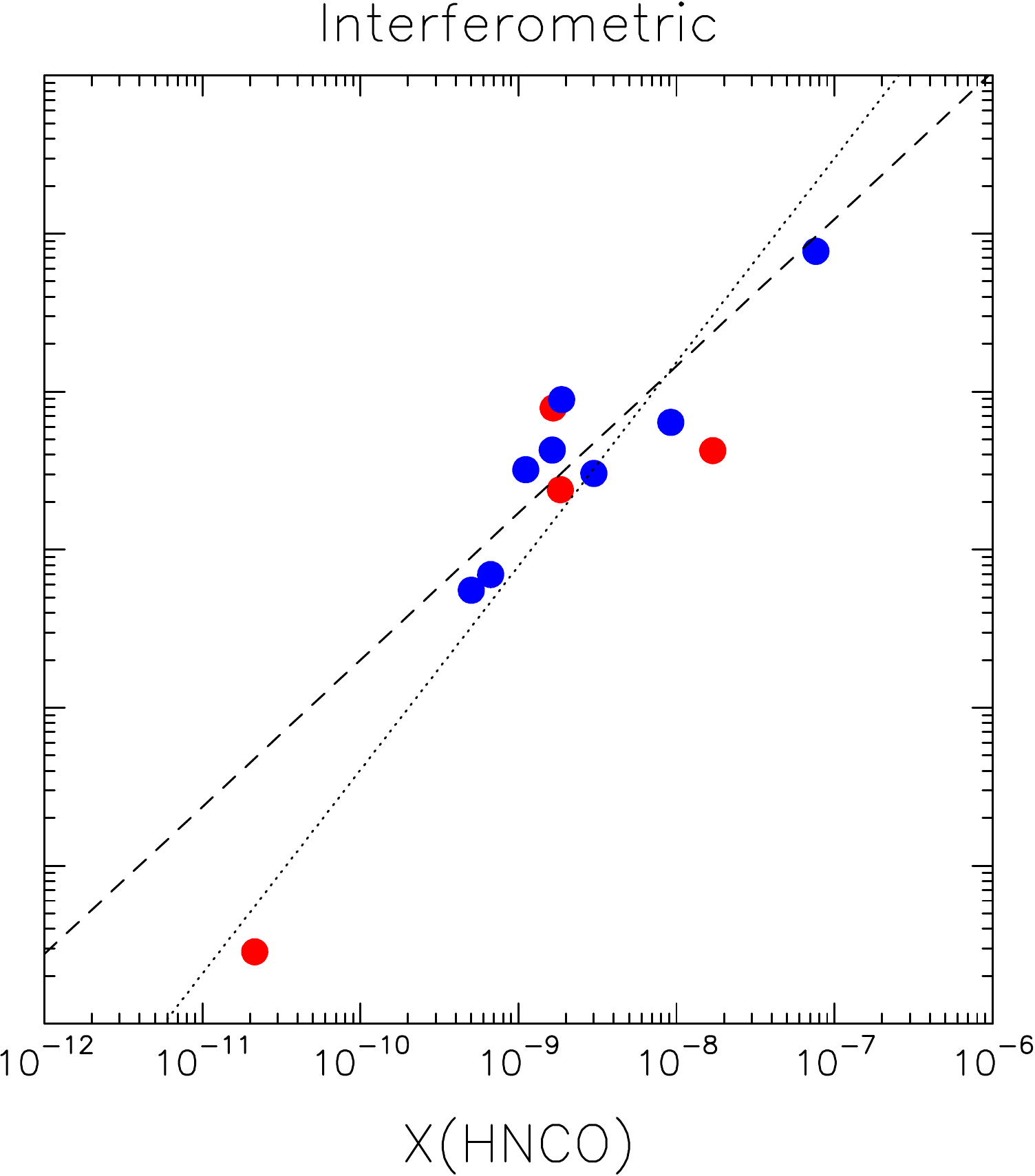}\\
    \end{tabular}
    \caption{Plot of NH$_2$CHO versus HNCO fractional abundances relative to H$_2$, colour-coded by mass category or type of object, derived from single-dish (\textit{left}) and interferometric (\textit{right}) observations reported so far in the literature. Filled and open circles depict compact, hot sources (i.e. hot cores/corinos), and more extended, cold protostellar envelopes, respectively. For clarity, low-mass sources for which formamide is not detected are marked with red crosses and arrows, which mark the 3$\sigma$ upper limit to its fractional abundance. The dashed and dotted lines correspond, respectively, to the best power-law fit derived by L\'opez-Sepulcre et al. \cite{ls15} considering only ``hot'' sources, and that derived by Quenard et al. \cite{quenard18a} including all sources regardless of whether or not they have detections of formamide.}
    \label{fcorr}
\end{figure}

One of the methods astronomers have recently used in order to shed light on the dominant formation pathway(s) of formamide is that of abundance correlations, i.e. by comparing the fractional abundance of formamide with that of other molecules suspected to be chemically linked. The molecule that has been most used for this comparison is isocyanic acid (HNCO), which contains two hydrogen atoms less than formamide. Figure\,\ref{fcorr} shows a plot of NH$_2$CHO against HNCO fractional abundances relative to molecular hydrogen, H$_2$, for all the sources in the literature with reported values of column densities for these two molecules, as well as total H$_2$ column density values\cite{gibb00,bisschop07,qin10,mendoza14,ls15,coutens16,feng16,imai16,allen17,ls17,oya17,marcelino18}. Single-dish and interferometric observations are separated into two plots. The plots also display the best power-law fit derived by L\'opez-Sepulcre et al. \cite{ls15}, $X$(NH$_2$CHO)\,=\,0.04\,$X$(HNCO)$^{0.93}$, and that derived by Quenard et al. \cite{quenard18a}, $X$(NH$_2$CHO)\,=\,32.14\,$X$(HNCO)$^{1.29}$. We caution that Quenard et al. included upper limits in their fit as if they were actual detections.

Figure\,\ref{fcorr} shows that there is a tight correlation between the two molecular species, which is almost linear and holds across several orders of magnitude in molecular abundance. This has led some authors to propose that HNCO and NH$_2$CHO are chemically linked, with either both of them forming from a common precursor, or one forming from the other\cite{bisschop07,mendoza14,ls15,Kanuchova2016}. Based on previous experimental work\cite{raunier04}, the most favoured formation route tended to be hydrogenation of HNCO on dust grains, followed by sublimation into the gas phase at sufficiently high temperatures, where it becomes detectable in relatively high amounts\cite{garrod08}.

However, these interpretations have been challenged by an experimental study that found hydrogenation of isocyanic acid in the solid phase to be unviable\cite{Noble2015} (see Sect.\,\ref{exp}), and alternative interpretations for the observed correlation between HNCO and NH$_2$CHO abundances have recently been proposed. In particular, Quenard et al. \cite{quenard18a} carried out a comprehensive study in which they modelled the published observational abundances of formamide with the aid of a gas-grain code, using in their chemical network all the known reactions involving molecules with amide bonds. They mainly considered three formation routes in their discussion: (i) hydrogenation of HNCO on grain surfaces; (ii) radical-radical formation on grains; and (iii) radical-radical association in the gas. Among other results (see Sect.\,\ref{theory}), they concluded that one should avoid over-interpreting correlations between the abundances of two molecules because it does not necessary imply that the two species are chemically linked. Indeed, in the case of HNCO and NH$_2$CHO, the authors found that the correlation is physical rather than chemical: the two species are formed through similar physical processes across different environments. Finally, Haupa et al \cite{haupa19} showed the possible linkage of HNCO and NH$_2$CHO in a solid phase experiment using a very cold H$_2$ matrix as a substrate (see Sect.\,\ref{solid}). In these conditions, these molecules are linked via a double cycle consisting of hydrogen extraction and the addition of hydrogen. This study demonstrated that formamide molecules are mostly converted to HNCO in the presence of a large amount of H atoms, which can explain why HNCO is systematically more abundant than formamide without necessarily being its parent molecule.

\subsubsection{Deuterated formamide}\label{deut}

Thanks to the superb sensitivity of ALMA, it has recently been possible to detect deuterated formamide and isocyanic acid, in the hot corino source IRAS\,16293--2422\cite{coutens16}. All three singly deuterated forms of NH$_2$CHO (NH$_2$CDO, cis- and trans-NHDCHO), as well as DNCO, were all found to co-exist in one of the two components of the protobinary (source B). 

The authors of this study argued that both HNCO and NH$_2$CHO are chemically linked based on their very similar deuteration fractions, measured to be 1\% and 2\%, respectively. From the significantly different deuteration fraction of 15\% estimated for formaldehyde (H$_2$CO), they ruled out the gas-phase reaction H$_2$CO\,+\,NH$_2$ $\longrightarrow$ NH$_2$CHO\,+\,H, which is favoured by the theoretical work by Barone et al. \cite{barone15} and previously suggested to be a dominant pathway leading to formamide in this particular source\cite{kahane13}. Instead, they considered grain-surface chemistry as the most likely mechanism to form both HNCO and NH$_2$CHO in IRAS\,16293--2422. As hydrogenation of HNCO was no longer considered, they proposed alternative surface reactions that could lead to these two molecules as products, such as NH\,+\,CO $\longrightarrow$\,HNCO and NH$_2$\,+\,H$_2$CO $\longrightarrow$ NH$_2$CHO + H\cite{fedoseev15,Fedoseev2016}.

Again, and analogously to what happens with abundance correlations (Sect.\,\ref{corr}), favouring one formation route or another based on similarities in molecular deuteration levels is not a secure method. Indeed, Skouteris and collaborators \cite{skouteris17} showed that the deuteration of the reactants is not simply transferred to the products of the gas-phase reaction. On the contrary, the deuteration ratios observed towards IRAS\,16293--2422 are in perfect agreement with those calculated for the H$_2$CO\,+\,NH$_2$ $\longrightarrow$ NH$_2$CHO\,+\,H gas-phase reaction. This is discussed in more detail in Sect.\,\ref{theory}.

\subsubsection{Formamide in outflow shocks}\label{shocks}

The observations of shocks produced by fast protostellar jets propagating through high-dense gas surrounding the newly born stars are unique laboratories where to investigate which formation mechanism is dominating for  iCOMs, and more specifically for formamide\cite{codella17,lefloch17}. Because of sputtering (gas-grain collisions) or shuttering  (grain-grain), atoms and molecules from both dust mantles and refractory cores are injected in the gas-phase, in turn triggering a rich warm-chemistry (e.g. Caselli \& Ceccarelli\cite{cc12}, and references therein). 

\begin{figure} 
    \centering
    \includegraphics[width=16cm]{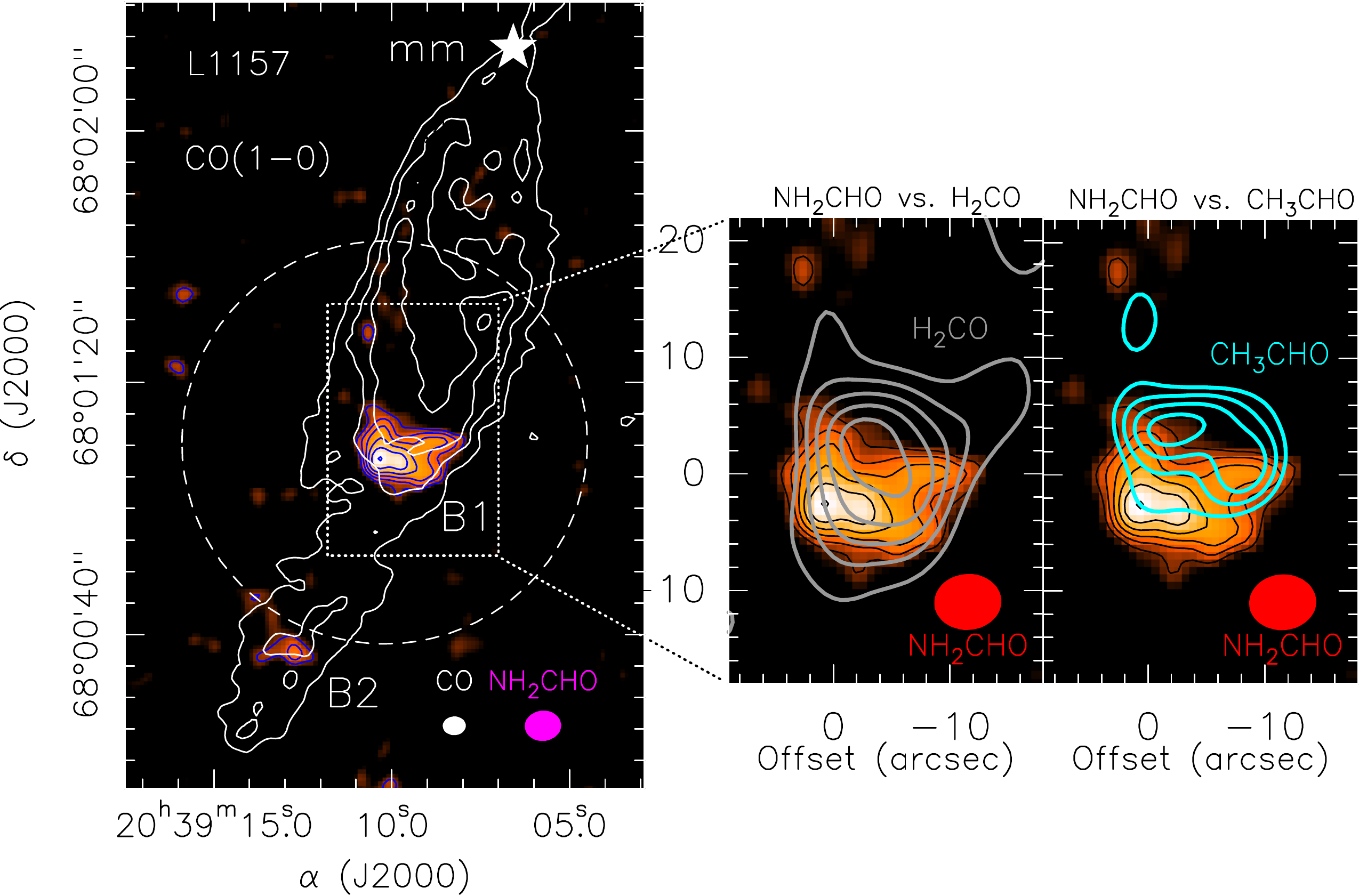}
    \caption{Adapted from Codella et al.\cite{codella17}. {\it Left panel:} The L1157 southern blue-shifted outflow cavities B1 and B2, here traced by the CO(1--0) emission (white contours; from Gueth et al. 1996), ejected by the L1157-mm protostar (white star). The colour scale is for the emission map of 
the NH$_2$CHO(4$_{\rm 1,4}$--3$_{\rm 1,3}$) line. The dashed circle shows the primary beam of the NH$_2$CHO image (64$''$). The ellipses in the bottom-right corner are for the synthesised beams of the NH$_2$CHO (magenta; $5.79'' \times 4.81''$) and CO (white; $3.65'' \times 2.96''$) observations, respectively. {\it Middle and right panels:} Chemical segregation within L1157-B1 (see also Benedettini et al. \cite{benedettini13} and Codella et al. \cite{codella15}). The NH$_2$CHO emission, associated with the B1 southern apex is compared with those of p-H$_2$CO(2$_{0,2}$--1$_{0,1}$), tracing then whole B1 structure (grey contours), and the CH$_3$CHO(7$_{0,7}$--6$_{0,6}$E+A), probing only the northern B1 portion (cyano). }
    \label{fshock}
\end{figure}

An illustrative case is represented by the L1157-B1 shock, located along the  chemically rich molecular outflow\cite{bachiller01} driven by the L1157-mm Class 0 protostar ($d$\,=\,250 pc). A precessing, episodic, and supersonic jet is launched by the L1157-mm protostar\cite{gueth96,podio16}. Two main blue-shifted cavities, called B1 and B2 (see Figure \ref{fshock}), have been excavated by the mass loss process. In particular, L1157-B1 consists of several shocks caused by different episodes of ejection impacting against the cavity wall. Within B1, two regions can be identified\cite{codella09}: (i) a northern arch-like structure associated with the youngest shock event as probed by the emission of dust mantle products such as HDCO\cite{fontani14}, and (ii) the apex, which is  the oldest one (kinematical age $\sim$ 1100 yr), being the farthest away from the protostar.

Several iCOMs, formamide among them, have been detected towards L1157-B1 using single-dishes\cite{arce08,mendoza14,lefloch17}. Interferometric surveys led also to image the iCOMs spatial distributions\cite{codella09,codella15}. In particular, in the context of the IRAM Large Program SOLIS (Seeds Of Life In Space)\cite{ceccarelli17}, formamide has been imaged with NOEMA (NOrthern Extended Millimetre Array): Figure\,\ref{fshock} (\textit{left}) shows that formamide is emitted from an extended region which is clearly associated with the B1 apex. More specifically, Codella et al. \cite{codella17} found a spatial anti-correlation (see Figure\,\ref{fshock} \textit{right}) between NH$_2$CHO and acetaldehyde (CH$_3$CHO): the NH$_2$CHO/CH$_3$COH abundance ratio increases moving from the northern part to the apex. Up-to-date astrochemical models\cite{skouteris17} demonstrated that the observed chemical differentation is possible only if the formamide observed in L1157-B1 is dominated by gas-phase chemistry and that the reaction NH$_2$\,+\,H$_2$CO $\to$ NH$_2$CHO\,+\,H explains the observations\cite{codella17}. Also the recent models of C-shock chemistry presented by Burkhardt et al. \cite{burkhardt19} confirm that the abundance of formamide is enhanced in the shocks through gas-phase reactions.

\subsection{Modern times: what are the next steps?}

The L1157-B1 shock case described above illustrates how obtaining a spatially-resolved image of formamide emission greatly helps to gain insight into the most likely formation routes leading to this molecular species, especially when the spatial segregation between different iCOMs can be explored\cite{codella17}. Coupled with suitable chemical modelling, this kind of observations can place valuable constraints on the formation mechanisms leading to formamide. 

However, while the detailed formamide study of the outflow shock region L1157-B1 represents a huge achievement towards understanding formamide synthesis in the ISM, it has some caveats. First, it concerns a region that is far from the actual protostellar disk where planets will eventually form. Indeed, the B1 shock is located about 20\,000\,au away from the central protostar L1157-mm (which, interestingly, does not show any detectable amounts of formamide emission), while a typical protostellar disk will have a size smaller than $\sim 100$\,au. Second, it is a shocked region, and therefore chemistry may work differently with respect to the less disrupted protostellar disk or inner envelope. Thus, for now we can conclude that gas-phase reactions are the dominant pathway leading to formamide in a protostellar shock.

What about the inner protostellar envelope, where material is expected to be subsequently delivered --whether re-processed or not-- onto the future planets?
Unfortunately, having a map of resolved emission of formamide and other iCOMs in these objects would probably not help. Indeed, contrarily to the shock region L1157-B1, where the shock passage causes a one-time injection of the grain mantles products into the gas-phase so that one can use the time-dependent composition to constrain the different formation routes, material continuously rains onto the inner protostellar envelopes so that no time-dependence effect can be used.

However, as mentioned above, the abundance ratio of the deuterated forms of formamide can provide a very strong constraint. In IRAS 16292--2422, we mentioned above that they favour the hypothesis of a gas-phase dominating formation route of formamide. To assess whether this is not just a case of a peculiar source, a systematic study of the formamide deuteration is needed. Given its low abundance, though, only the high sensitivity observations of the new facilities such as ALMA and IRAM-NOEMA will be able to provide useful detections in more sources.

\subsection{Formamide in other environments}

While the present review focuses mostly on understanding the chemistry of formamide in regions of star formation, it is worth mentioning that this molecule has been observed in other kinds of environments (see Table\,\ref{tsou}). These include (i) translucent cloud material along the line-of-sight towards Sgr\,B2, where formamide has been detected in absorption\cite{corby15,thiel19}; (ii) diffuse clouds in a spiral galaxy\cite{muller13}; and (iii) a disk of material surrounding a white dwarf-brown dwarf merger thought to have occurred in 1670\cite{eyres18}. The two first sources are similar in that they both concern diffuse or translucent molecular clouds, which is an interesting clue to the formation mechanism of NH$_2$CHO, as gas-phase chemical reaction networks are believed to dominate in these environments\cite{turner00}.

\section{The chemistry of interstellar formamide: experiments}\label{exp}

There are two main processes thought to be efficient in governing the chemistry of interstellar formamide: (i) gas-phase reactions, and (ii) grain surface chemistry. Both experimental and theoretical work need to be continuously updated in order to feed the network of chemical reactions employed by the chemical models that are used to interpret the observations. This section summarises the various experiments carried out that have yielded formamide among their products, both on the solid state, and in the gas-phase, while theory will be the subject of Sect.\,\ref{theory}.


\subsection{Experiments on solid state formamide}\label{solid}

 In a simplified way, the formation of iCOMs on interstellar dust grains that leads to their observation in the gas phase can be summarily divided into three phases: i) the growth of the molecular mantle of about 100 layers thick on the grains through the accretion of the gas phase of chemically generally simpler species (e.g. Taquet et al.\cite{taquet12}). The emblematic example is the formation of methanol by successive hydrogenation of CO (CO$_g$ +n H$_g$ $\longrightarrow$ CH$_3$OH$_g$)
(ii) Ice processing, by electronic, photonic (UV) or energetic particle (ions) impact
(iii) Desorption of the mantle which may induce its chemical transformation, (e.g. NH$_3$ and CO$_2$ can react \citep{Theule2013}).
Of course the three steps are intertwined, one not stopping to give way to another.

Historically, it is the photolysis of prototype ice mixtures containing the elements H, N, C, and O, such as NH$_3$-CO, that has identified the presence of formamide \citep{Grim1989}.
The presence of formamide was confirmed from other mixtures or energetic processing.  Energetic electron bombardment of CO--NH$_3$ ice mixtures\cite{jones11} or other mixtures\cite{bergantini14,hg15}, ion irradiation of H$_2$O--HCN ices at 18\,K\cite{gerakines04},  warm-up of photolysed ice mixtures of H$_2$O, CH$_3$OH, CO, and NH$_3$\cite{bernstein95}, lead to the formation of formamide. UV photolysis and proton irradiation of HCN containing ices\cite{gerakines04} or by UV irradiation of pure HNCO ice\cite{raunier04} or mixed HNCO:CH$_4$\cite{Ligterink2018} or H$_2$O, CO, NH$_3$ ice mixtures\cite{demyk98,VanBroekhuizen2004}, all produces some formamide. But in each case formamide was accompanied by many other molecular species such as HNCO and OCN$^-$ or even CH$_3$NCO, among others. 

Kanuchova et al. \cite{Kanuchova2016} have shown that the irradiation with protons or helium ions of different solid mixtures containing water, methane, ammonia, or nitrogen at low temperatures drives the synthesis of formamide together with isocyanic acid (HNCO). By studying different mixtures and changing the proportions of nitrogen atoms in their mixtures subjected to ion bombardment, they showed that the HNCO/NH$_2$CHO ratio was almost always less than 1, disagreeing with most observations and suggesting the possibility of a distinct formation pathway (in the gas phase) of HNCO. On the contrary, in their study of UV photolysis of a mixture of NH$_2$OH (formed in situ) and CO or H$_2$CO, Fedoseev et al. \cite{Fedoseev2016} seemed to point to a common chemical history. Indeed, photolysis of their dynamically formed ice mixtures produce NH$_2$ radicals (from the decomposition of NH$_2$OH) and, as in the gas phase, NH$_2$ can react either with CO, HCO or H$_2$CO, the former giving birth to HNCO and the later to NH$_2$CHO. Therefore, following these authors, one could attribute the common origin of NH$_2$CHO and HNCO to the co-existance of the various degrees of hydrogenation of CO and the presence of the NH$_2$ radical.

Despite the diversity of these experiments and their results, which are sometimes difficult to compare or reconcile, it appears that formamide is easily produced in the solid phase by energy input into molecular ice containing H N C O precursors, even if formamide is accompanied (or even diluted) by many other chemical compounds.

 \begin{figure} 
    \centering
    \includegraphics[width=12cm]{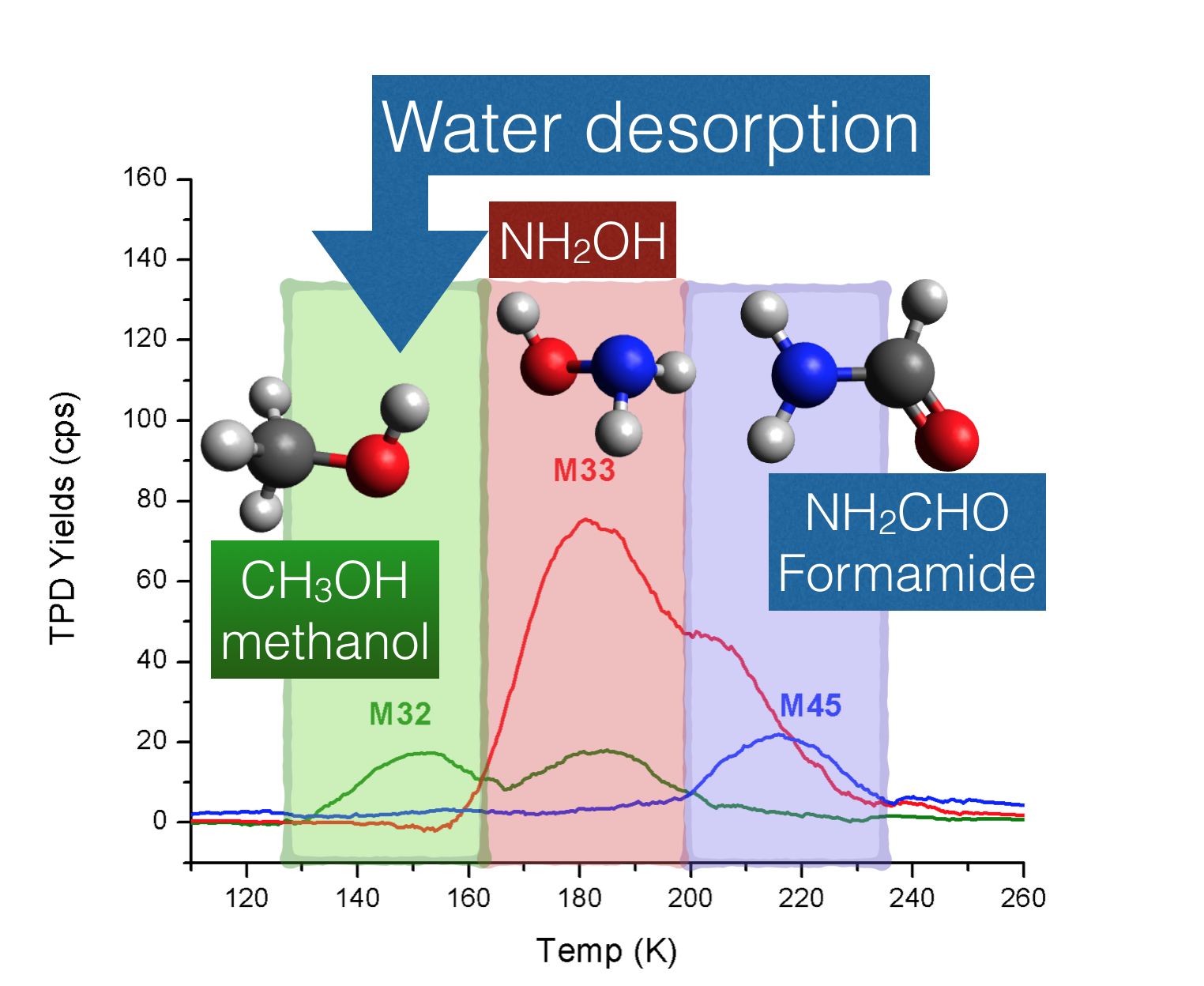}
    \caption{Thermally programmed desorption (heating ramp 0.2K/s) profiles (green, red and blue solid lines) of a condensed mixture of a few layers of water, methanol, hydroxylamine and formamide  formed at 10 K on a gold plate (adapted from Dulieu et al. \cite{Dulieu2019}). Under these experimental conditions water and methanol are co-desorbing at around 150\,K (green area), hydroxylamine (NH$_2$OH) at around 180\,K while formamide is desorbing above 200\,K (blue area).}
    \label{Fig:formamidedesorption}
\end{figure}

This aspect has an impact on the potential return of the molecule to the gas phase. On this point, the experiences converge well. Formamide is a species that is refractory to the sublimation of the water ice mantle \citep{Dawley2014,Urso2017,Chaabouni2018}, and its desorption only occurs around 220\,K under laboratory conditions whereas the mixed water is desorbing at about 160\,K. The binding energy distribution of formamide was measured and compared (see Table 1 in Chaabouni et al.\cite{Chaabouni2018}). But even more so, when other organic residues are produced (case of energetic processing) or the material is porous (nanoporous silicates), sublimation can be greatly delayed and sublimation can be observed at much higher temperatures (400\,K), which is a very interesting aspect for comet science, or the delivery of materials to planets' atmosphere. Of all the iCOMs commonly detected in ISM, formamide is particularly refractory. Figure~\ref{Fig:formamidedesorption} illustrates this aspect. It shows thermal desorption profiles of a mixture of water, methnaol, hydroxilamine (NH$_2$OH) and formamide. The temperature profiles (green, red, blue lines) of iCOMs are displayed and formamide appears to desorb at the highest temperature, indicating the highest binding energy to the surface (a gold plate in this example).

Regarding the construction of the ice mantle on grains, fewer studies have been published, but the idea that formamide could be the product of hydrogenation of HNCO has been ruled out \citep{Noble2015}. However, Haupa et al. \cite{haupa19} recently showed the possible link between HNCO and NH$_2$CHO in the solid phase using a very cold H$_2$ matrix (3\,K) as a substrate, in the presence of H atoms. In these conditions, HNCO and NH$_2$CHO are linked via a double cycle consisting of H-abstraction and H-addition. Moreover, it is possible to obtain a large fraction of formamide by hydrogenation of a mixture of H$_2$CO and NO \citep{Dulieu2019}, without formation of HNCO. This chemical pathway, only efficient if NO and H$_2$CO are neighbours on the surface of dust grains, seems to be sufficient to produce about the same ratio as that observed in ISM (compared to methanol or H$_2$).

In conclusion, in the solid state, the formation of HNCO and NH$_2$CHO may be linked by direct hydrogenation or H-abstraction. These two molecules can also be linked by their common precursors, i.e. the different degrees of hydrogenation of CO (H$_2$CO in particular) and another radical containing N, such as NH$_2$, NHOH, or even N atoms, although no experiment has yet been carried out with the latter.

\begin{figure} 
    \centering
    \includegraphics[width=14cm]{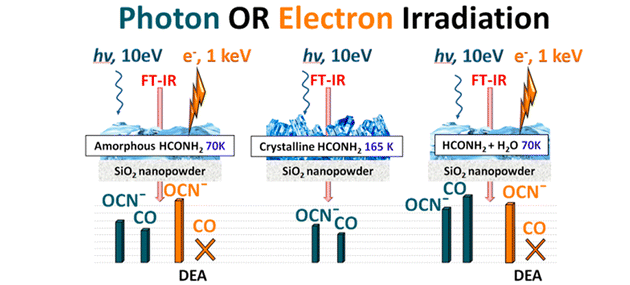}
    \caption{Products that result from processing of formamide on icy mantles by electrons and/or UV photons. Reprinted with permission from Dawley et al. \cite{Dawley2014b}, Copyright 2014 The Journal of Physical Chemistry A.}
    \label{Fig:formamideprocessing}
\end{figure}

There is another aspect of the experimental work that studies the routes of destruction, or the stability of formamide, which is sometimes omitted in astrochemistry. Electron bombardment or UV photolysis have been studied by Dawley et al. \cite{Dawley2014b}. The major observable destruction channel lead to OCN$^-$, and CO is also observed. Figure\,\ref{Fig:formamideprocessing} summarises the main findings of this study.

Brucato et al. \cite{Brucato2006} studied the impact of ions on formamide films. They estimated that 64$\%$ of the formamide molecules are destroyed at 12 eV/16 amu (a low dose for such experiment). Extrapolating to the irradiation dose expected in the dense ISM, they inferred that about 20$\%$ of the frozen formamide molecules are able to survive in dense medium on a time scale of 10$^8$ years. More interestingly, they stated that ``the species produced after irradiation of formamide are mostly the same as those produced in a large number of irradiation experiments conducted by different groups on different icy mixtures containing simple H, O, C, and N bearing molecules\cite{grim87,demyk98,moore03,hudson01,palumbo00,palumbo04}." In addition, they pointed out the role the ammonium cyanate molecule: ``In fact the CN stretching in OCN$^-$ is considered responsible for the 2165 cm$^{-1}$ band observed in almost all the observations of molecular clouds and in young stellar objects." Finally, a very recent experimental study demonstrated that formamide is easily converted to HNCO by the reactions with thermalized H atoms in a cold para-H2 matrix kept at ca. 3 K. In the same study, it was actually demonstrated that formamide molecules are mostly converted to HNCO in the presence of a large amount of H atoms, which can explain why HNCO is systematically more abundant than formamide without being its parent molecule. 

As a current understanding of the formation and evolution of formamide on solid state, it is clear that it is a refractory molecule (in comparison with water and many other iCOMs). It may be formed by many different chemical pathways from different initial ice mixtures containing H, N, C, and O, that are, in the case of energetic pathways, also destructive and produce a lot of companion molecules, the most detectable being identified as OCN$^-$, probably associated with NH$_4^+$. Finally, it has to be noted that all the mentioned experiments do not fully reproduce the ISM conditions, but rather have to be considered to understand the processes possibly at work on the surfaces of interstellar grains. Indeed, all the irradiation experiments use doses that are several orders of magnitude larger than those in the ISM, for practical reasons of sensitivity of the experiments (after all, ISM objects have millions years at their disposal while we only have a lifetime at most!). On the same vein, the experiments using H$_2$CO and NO to synthesise formamide without extra source of energy, are efficient only when the two species are next to each other. In general, this is highly improbable, as the major interstellar ice component is water, so that the probability that H$_2$CO and NO are next to each other is very tiny (see e.g. Ceccarelli et al. \cite{ceccarelli18} for the example of similarly abundant species on the grain mantles).

\subsection{Experiments on gas-phase formamide}\label{gas}

In the cold ISM, ion-neutral reactions have been expected to dominate the chemistry of iCOMs for decades. Only recently, it has become clear that neutral-neutral reactions can also be important in the ISM\cite{shannon13,balucani15,skouteris18}. In the case of formamide, we already mentioned the reaction NH$_2$ and H$_2$CO for which no laboratory experiments exist to our best knowledge.
On the contrary, several experimental studies have been investigating the formation and destruction routes of NH$_2$CHO from ion-neutral reactions. 

The most abundant N-bearing molecular ions in the ISM are NH$^+$, NH$_2^+$ and CN$^+$. On the other hand NH$_4^+$ is a dead end which locks the N atom, as NH$_3$ has a strong proton affinity and does not transfer easily its proton to any other molecules.

The first step is to create an amide bond starting from the two most abundant molecular ions and from the most abundant neutrals.

Adams et al. \cite{adams80} investigated the reactions of NH$_{n}^+$ ions ($n\,=\,0$ to 4) with CO and OCS (br\,=\,branching ratio):
\begin{eqnarray*}
NH^+ + CO &\rightarrow NCO^+ + H ~~~~~ (br 55\%, T = 300\,K)\\
NH_2^+ + OCS &\rightarrow H_2NCO^+ + S ~~~~~ (br 15\%, T = 300\,K)
\end{eqnarray*}
Note that NH$_2^+$ + CO$_2$ gives no reaction.

Several experimental studies address the reactivity of the cyano radical cation CN$^+$ with several neutrals (H$_2$O, O$_2$, CO$_2$, NO and N$_2$O), which gives the NCO$^+$ ion or its hydrogenated form, the isocyanic acid ion HNCO$^+$\cite{mcewan83,raksit84,anicich93}:
\begin{eqnarray*}
CN^+ + H_2O \rightarrow HNCO^+ + H ~~~~~ (br 20\%, T = 298\,K)\\
CN^+ + O_2 \rightarrow NCO^+ + O ~~~~~ (br 20\%, T = 298\,K)\\
CN^+ + CO_2 \rightarrow NCO^+ + CO ~~~~~ (br 30\%,T = 298\,K)\\
CN^+ + NO \rightarrow NCO^+ + N ~~~~~ (br 25\%, T = 298\,K)\\
CN^+ + N_2O \rightarrow NCO^+ + N_2 ~~~~~ (br 20\%, T = 298\,K)
\end{eqnarray*}

Ion-neutral reactions starting from the hydrogenated form of the cyano radical cation, namely HCN$^+$ and HNC$^+$, give also the isocyanic acid cation HNCO$^+$\cite{petrie90}:
\begin{eqnarray*}
HNC^+ + O_2 \rightarrow HNCO^+ + O ~~~~~ (br 75\%, T = 300\,K)\\
HNC^+ + N_2O \rightarrow HNCO^+ + N_2 ~~~~~ (br 55\%, T = 300\,K)
\end{eqnarray*}

In addition to cyanides, the nitric oxyde ion can react with carbon chains to form an amide bond\cite{midey01}:
\begin{equation*}
NO^+ + l-C_4H_8 \rightarrow HNCO~or~H_2NCO~or~NH_2CHO ~~~~~ (br~small, T = 225\,K)
\end{equation*}
Note that the NO$^+$ + l-C$_4$H$_8$ gives a little bit of NH$_2$CHO.

Once the amide bond is formed, the second step consists of protonating or hydrogenating the amide bond bearing molecule.

Studies show that the H$_2$NCO$^+$ radical ion can be produced from isocyanic acid, in its neutral and ionized form, HNCO or HNCO$^+$\cite{wight80}:
\begin{eqnarray*}
HCO^+ + HNCO \rightarrow H_2NCO^+ + CO ~~~~~ (br 100\%, T = 300\,K)\\
NCO^+ + HNCO \rightarrow HNCO^+ + NCO ~~~~~ (br 100\%, T = 300\,K)\\
HNCO^+ + HNCO \rightarrow H_2CO^+ + NCO ~~~~~ (br 100\%, T = 300\,K)
\end{eqnarray*}

The missing step is the hydrogenation of the radical cation NH$_2$NCO$^+$ or the protonation of the radical H$_2$NCO to get NH$_2$COH$^+$. Although there are no experimental studies on that, abundant ions such as H$_3^+$ or HCO$^+$ should ensure the proton transfer.

The reaction of NH$_2$COH$^+$ (protonated NH$_2$CO radical) with alkenes gives protonated formamide NH$_2$CHOH$^+$\cite{bouchoux02}:
\begin{align*}
NH_2COH^+ + C_2 H_4 &\rightarrow& NH_2CHOH^+ + C_2H_3 ~~~~~(, T = 300\,K)\\
&\rightarrow& NH_2CO^+ + C_2H_5  ~~~~~(, T = 300\,K)\\
&\rightarrow& C_2H_5^+ +NH_2CO  ~~~~~(, T = 300\,K)\\
\end{align*}

Other reactions, NH$_2$COH$^+$ + C$_3$H$_6$, NH$_2$COH$^+$ + C$_4$H$_8$\cite{bouchoux02}, give protonated NH$_2$CHOH$^+$ as a minor product. Even though there are no experimental studies, it should be fairly easy to deprotonate formamide using a strong proton acceptor (such as NH$_3$), which will give:
\begin{equation*}
NH_2CHOH^+ + NH_3 \rightarrow H_2NCHO + NH_4^+ ~~~~~ (br:?, T = 300\,K)
\end{equation*}

Dissociative electron-ion recombination is also possible, but it is expected to cause a fission of the amide bond with only a small percentage of neutral formamide being formed (unpublished results).

Aside from these formation routes on the way to formamide, some reactions can break the amide bond\cite{bouchoux02}:
\begin{equation*}
NH_2CHO^+ + H_2O \rightarrow NH_3(H_2O)^+ + CO  ~~~~~(T = 300\,K)
\end{equation*}

In summary, we know how to form an amide bond from abundant interstellar molecules. To go to formamide, two steps are not known,
(i) the formation of the protonated radical NH$_2$COH$^+$, and (ii) the deprotonation of protonated formamide. While the latter step should be fairly easy with a strong proton acceptor such as NH$_3$, the former step needs to be studied carefully.
%

\section{The chemistry of interstellar formamide: theory}\label{theory}

\subsection{Theory on gas phase reactions}

As already noted, a possible formation route of formamide in the gas phase involves the reaction between formaldehyde, a widely spread molecule in the ISM, and the amidogen radical, NH$_2$, which can be formed starting from ammonia or by the partial protonation/hydrogenation of atomic nitrogen. This reaction was first considered in the OSU data base\cite{garrod08}. No experimental nor theoretical data were available on that bimolecular reaction back then. Therefore, under the assumption that it is a barrier-less reaction because it involves an open shell radical, a rate coefficient in the gas kinetics range ($10^{-10}$\,cm$^3$\,s$^{-1}$) was suggested. Later on, however, Garrod\cite{garrod13} disregarded such an assumption after the theoretical study by Li \& Lu\cite{li02} focused on the product channel leading to NH$_3$ + HCO. Even though Li \& Lu\cite{li02} did not characterise the channel leading to NH$_2$CHO + H, after a comparison with the analogous OH + H$_2$CO reaction, Garrod (2013) concluded that the role of the NH$_2$ + H$_2$CO $\rightarrow$ NH$_2$CHO + H was irrelevant. The reaction was therefore excluded from the gas-phase network and only considered as a possible formation route of formamide when occurring on the icy mantles of interstellar grains\cite{garrod08,garrod13}.

As a matter of fact, the reaction NH$_2$ + H$_2$CO is challenging to investigate in laboratory experiments because it is difficult to generate a free radical like NH$_2$ in a controlled manner and with sufficient number density in the presence of H$_2$CO, another tricky gaseous species to generate in laboratory experiments. Urged by the possible importance of this reaction and the lack of experimental data, Barone et al.\cite{barone15} performed the first electronic structure calculations of the potential energy surface of that reactive system at the B2PLYPD3/m-aug-cc-pVTZ and CBS-QB3 levels of theory and then provided an estimate of the rate coefficient by combining capture theory and RRKM calculations. The main findings of this first theoretical study can so be summarised: the interaction of the NH$_2$ radical with the $\pi$ system of formaldehyde leads to a bound addition intermediate, H$_2$CONH$_2$, that, in turn, fragments into the products NH$_2$CHO and H by the fission of one of the two C--H bonds. Remarkably, along the minimum energy path there is a transition state in the exit channel leading to the products which lies at an energy level comparable to that of the reactants. At the level of calculations employed by Barone et al.\cite{barone15}, the energy level of the transition state is slightly above the energy level of the reactants (+0.2\,kJ/mol when considering the zero point energy, ZPE, of all species), but this does not prevent the reaction from being very fast because of tunnelling through the barrier. The competition between tunnelling through the exit barrier and back dissociation to reactants causes a strong temperature dependence of the rate coefficients which reduces by one order of magnitude between 10 and 100 K. A numerical fit of the rate coefficient provided the best fit parameters $\alpha$, $\beta$ and $\gamma$ to be included in astrochemical models according to the modified Arrhenius formalism $k(T) = \alpha \times (T/300)^\beta exp[-\gamma/T]$. To be noted that $\alpha$, $\beta$ and $\gamma$ so provided do not have any physical meaning as they are only the numerical parameters that best represent the trend of the calculated $k$ as a function of temperature.

Later on, Song and K\"astner\cite{song16} performed calculations at M06-2X80/def2-TZVP27 level using NWCHEM 6.681 and single-point energies and vibrational frequencies on the UCCSD(T)-F1276,77/cc-pVTZ-F1278 level. They derived a much higher barrier located in the entrance channel and estimated a rate coefficient of only $5.3 \times 10^{-22}$\,cm$^3$\,s$^{-1}$ at 100\,K which makes the gas-phase NH$_2$ + H$_2$CO reaction irrelevant in the conditions of interstellar clouds. After that, Vazart et al.\cite{vazart16} performed a thorough exploration of the complete potential energy surface of NH$_4$CO, with the main body of calculations performed at the B2PLYP-D3 level and the energy for critical points calculated at the CCSD(T)/CBS+CV+fT+FQ level (see below). According to their description, the reaction starts with the formation of one of two possible van der Waals complexes (in one case, one of the H atoms of NH$_2$ is interacting with the oxygen atom of formaldehyde; in the second case, the nitrogen atom of NH$_2$ interacts with the carbon atom of formaldehyde (see Figure\,1 of Skouteris et al.\cite{skouteris17}) without any energy barrier. Once one of the two van der Waals complexes is formed, however, the system needs to overcome a transition state (vdW TS in Figure 1 of Skouteris et al.\cite{skouteris17}) connecting both wells of the complexes to the addition bound intermediate H$_2$NH$_2$CO. As already found by Barone et al., the addition intermediate needs to overcome a second transition state where one of the C--H bonds is weakening before breaking up with the formation of NH$_2$CHO and a free H atom. Therefore, the peculiarity of this system is that both transition states are characterised by an energy level which is very close to that of the reactants. For this reason, it is critical to use the best possible level of theory as the typical uncertainty of quantum calculations is within $\pm\,4$ kJ/mol, that is enough to make the transition states either emerged (above) or submerged (below the energy of the reactants). This is dramatically important in the context of cold interstellar objects because of the extremely low temperatures of interest. For this reason, Vazart et al.\cite{vazart16} and Skouteris et al.\cite{skouteris17} performed additional calculations based on coupled cluster method. In particular, the coupled-cluster singles and doubles approximation augmented by a perturbative treatment of triple excitations (CCSD(T)) has been employed in conjunction with extrapolation to the complete basis set limit and inclusion of core-correlation effects (CCSD(T)/CBS+CV). At this level of calculations, the energy level of the van der Waals transition state is still slightly above the energy of the reactants’ asymptote, but when a full treatment of triple (fT) and quadruple excitations (fQ) is used, the energy level of this transition state decreases from 3.6 kJ/mol to 2.05 kJ/mol with full-T and to 1.67 kJ/mol with full-T and full-Q. CCSD(T)/CBS+CV+fT+fQ is, therefore, the highest level of calculations ever employed for this reactive system. Not only that: by extrapolating to the full configuration interaction limit, the electronic energy of this transition state drops slightly below the reactant level\cite{vazart16}, in line with the variational principle of approximating the energy from above. In addition to that, Vazart et al. and Skouteirs et al. claimed that the use of standard methods to evaluate the ZPE correction for the van der Waals complex and its transition state is not warranted because the three new vibrational modes in the van der Waals complex are a loose stretching mode, the frequency of which will be largely overestimated even with a perturbative approach including anharmonicity, and two loose bending modes that are almost free rotations. Because of the risk of largely overestimating the ZPE effect for the van der Waals complex and its transition state, Vazart et al. and Skouteris et al. concluded that it is better to omit it. Nevertheless, the higher level of calculations employed affected also the results of kinetic calculations, and a smaller rate coefficient for all values of temperature was derived. The new alpha, beta and gamma parameters are, respectively, $7.29 \times 10^{-16}$, --2.56 and 4.88. Once again, these parameters originate from the best fit of the calculated rate coefficients and do not hold any specific physical meaning per se. In particular, the value of gamma does not represent the reaction barrier. 

\begin{figure} 
    \centering
    \includegraphics[width=10cm]{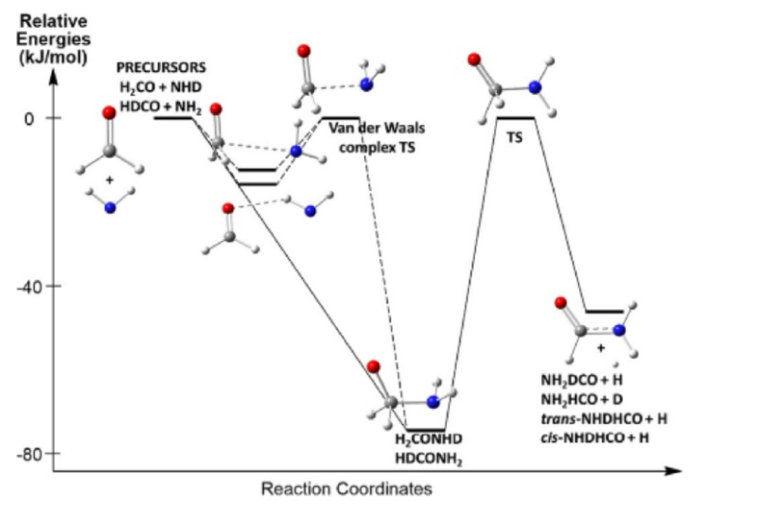}
    \caption{Path of the gas-phase reactions of deuterated formaldehyde with amidogen (H$_2$CO + NHD $\rightarrow$ NHDHCO + H and HDCO + NH$_2 \rightarrow$ NH$_2$DCO + H) showing that the two barriers are submerged. Figure adapted from Skouteris et al. \cite{skouteris17}.}
    \label{ftheory1}
\end{figure}

The rate coefficient derived by Barone et al.\cite{barone15} was employed to simulate the formamide production in two different objects, the shock region L1157-B1 and the protostar IRAS16293-2422, both mentioned in the Section 2. The agreement between the observed values and the computed ones were extremely encouraging and supports a gas-phase origin of formamide in these two objects. The new values derived in Skouteris et al.\cite{skouteris17} were instead used to simulate the formamide spatial distribution in L1157-B1\cite{codella17} (see Sect.\,\ref{shocks}) and provided a very strong constraint on the gas-phase origin of formamide there. Also, Skouteris et al. characterized the isotopic effect when considering partially deuterated reactants in an attempt to simulate the D-enrichment of formamide in the protostar IRAS\,16293--2422 as it resulted from an ALMA detection campaign\cite{coutens16}. Also in this case, the excellent agreement between the observed and predicted values is in favour of a gas-phase origin of formamide, as also discussed in Sect.\,\ref{deut}.

For completeness, we mention here the modelling study by Quenard et al.\cite{quenard18a} that, instead, found that the NH$_2$ + H$_2$CO gas phase reaction is not efficient enough to explain the observed abundance of formamide. However, the authors did not use either the rate parameters derived by Barone et al.\cite{barone15} or those derived by Skouteris et al.\cite{skouteris17} but a combination of the two. In particular, they adopted the $\gamma$ value from the former and the $\alpha$ and $\beta$ values from the latter, thus producing another $k(T)$ which does not resemble none of the two. After that, Quenard et al., indeed, commented on the effect of the barrier identified with the $\gamma$ value, but actually the $\gamma$ parameter has lost its original meaning of the modified Arrhenius equation, being only a best-fit parameter. Notably, the rate coefficient derived by Skouteris et al. is systematically smaller than that derived by Barone et al. in the entire range of temperature investigated.

Several ion-molecule reactions have also been considered from a theoretical point of view\cite{redondo14a,redondo14b,spezia16}. In particular, reactions involving ionic or protonated forms of nitrogen compounds (NH$^{+}_3$ , NH$^+_4$ , NH$_3$OH$^+$, and NH$_2$OH$^+$) and neutral molecules having one carbonyl group (H$_2$CO and HCOOH), as well as those involving protonated formaldehyde (H$_2$COH$^+$) with NH$_2$OH or NH$_3$ have been considered. None of them was found to be a possible formation route of interstellar formamide, either because they form other products or because the reaction profile is characterized by high energy barriers.

\subsection{Theory for ice-assisted reactions}

Song and K\"astner\cite{song16} used a theoretical approach to shed light on the hydrogenation process of HNCO on amorphous solid water surfaces, a process which was long thought to be responsible for formamide formation before the experimental work of Noble et al.\cite{Noble2015} and Fedoseev et al.\cite{Fedoseev2016}. More specifically, they constructed a model of interstellar amorphous solid water surface by classical molecular dynamics simulations and considered 113 possible orientations of HNCO to derive all possible binding energies. The values of possible binding energies varied between 0 and 100\,kJ/mol, but the barrier for the hydrogenation reaction resulted to be essentially independent of the binding energy. After optimising the structure, they calculated tunnelling rate constants for the H + HNCO $\rightarrow$ NH$_2$CO reaction, either in the gas phase ($T\,=\,95$--289\,K) and on the amorphous solid water at temperatures as low as 100\,K by using the instanton theory. The activation barrier for the surface reaction was found to be 4.4\,kJ/mol lower than in the gas phase, thus confirming that the interaction with the water molecules of the ice does not significantly change the barrier of reactive systems with respect to the gas-phase case, as already pointed out in previous work (see for instance Rimola et al.\cite{rimola14}). In addition to that, this barrier height reduction was not seen to imply a larger rate coefficient below 240\,K. This is due to the fact that also the width of the barrier has changed moving from the gas phase reaction to that catalysed by amorphous solid water. The effect of the variation of the width reduced the extent of tunnelling rate for this system. In both cases, tunnelling was seen to accelerate the hydrogenation reaction by many orders of magnitude at low temperature. For instance, at a temperature of $\sim\,100$\,K, the inclusion of tunnelling increased the rate coefficient by a factor $10^{10}$ and $10^8$ for the gas-phase and the surface reactions, respectively. Because of the important role played by tunnelling, a strong kinetic isotopic effect is expected when comparing the rate coefficient of H + HNCO $\rightarrow$ H$_2$NCO to that of D + HNCO $\rightarrow$ NHDCO both in the gas phase and on the amorphous solid water surface. This effect was quantified by Song and Kastner\cite{song16}: in the gas phase, the reaction involving H is faster than that involving D by a factor of 231, which is slightly reduced to 146 in the case of the ice-assisted reaction. This aspect is extremely important because it might affect the degree of deuteration of formamide with respect to HNCO. We recall that a deuteration ratio of ca. 1\% for HNCO and 2\% for NH$_2$CHO was determined for IRAS\,16923--2422\cite{coutens16}, so that the hydrogenation of HNCO on interstellar ice to form formamide can be ruled out (see Sect.\,\ref{deut}).

A fairly comprehensive theoretical work on the possibility of forming formamide on ice is the one provided by Rimola et al.\cite{rimola18} and Enrique-Romero et al. (this volume). The former characterised different processes, all possibly leading to formamide, by simulating their occurrence on a cluster of 33 H$_2$O molecules chosen to represent the surface of amorphous interstellar water ice. All the calculations were single-point energy calculations at BHLYP/6-311++G(d,p) onto the optimized BHLYP/6-31+G(d,p) geometries and ZPE corrections are those provided by the frequency calculations at BHLYP/6-31+G(d,p). They first started by considering the process proposed by Garrod et al.\cite{garrod08}, that is the radical-radical association reaction of NH$_2$ + HCO on the surface of water ice via a Langmuir-Hinshelwood mechanism. The calculations show that formamide can indeed be formed, but in competition with formation of NH$_3$ and CO through a direct H transfer process. The final outcome of the NH$_2$ + HCO reactivity depends on the relative orientation of the two radicals on the ice surface: only when the C atom of HCO points toward the nitrogen atom of NH$_2$ the formation of the bond is possible, by overcoming a barrier of 3\,kJ/mol (to be noted that in a case like this, the tunnel effect cannot help to overcome the barrier). In addition to that, a large binding energy for both species was derived (2103\,K for HCO and 4026\,K for NH$_2$) which makes it difficult for them to move along the ice surface.

With their theoretical method, Rimola et al. also explored two other processes, never considered before, namely the reaction of either HCN or CN with water molecules of the ice mantle. The reaction with HCN has been found to be characterised by large energy barriers and, therefore, its contribution is negligible under the conditions of interstellar ice. However, the reaction with the CN radical can occur, possibly leading through multiple steps to the formation of NH$_2$CHO. For this reaction, water molecules of the ice act as catalytic active sites since they help the H transfers involved in the process, thus reducing the energy barriers (compared to the gas-phase analogous reaction). Additionally, Rimola et al. applied a statistical model to estimate the reaction rate coefficient when considering the cluster of 33 H$_2$O molecules as an isolated moiety with respect to the surrounding environment (i.e., the rest of the ice). Their conclusion is that CN quickly reacts with a molecule of amorphous ice and that it can synthesise formamide, even though the efficiency of the NH$_2$CHO formation is difficult to estimate as it depends on the unknown number of ice water active sites and the fine details of energy transfer through the ice body itself. This has two important consequences on the modelling of interstellar surface chemistry. First, the H$_2$O molecules of the ice, usually considered as an inert support in astrochemical models, can instead react with active radicals, like CN, forming more complex species, and can also act as catalysts by helping H transfer processes. Second, most of the involved intermediate steps toward formamide formation on the 33-H$_2$O molecule cluster are quite fast, so it is unlikely that the energy released in each of them can be dispersed in the entire ice body of the grain. In other words, the system does not necessarily equilibrate at the grain temperature in each intermediate step, as assumed in all current models, and the localised energy can promote endothermic or high barrier processes in small portions of the ice before complete equilibration.

The time scale of energy redistribution within the ice molecules, a poorly characterised process, should be explicitly accounted for if a realistic model of grain surface chemistry is pursued.

\section{Final discussion and conclusions: the full picture}\label{final}

Motivated by the multiple studies that highlight the potential of formamide as a key pre-biotic precursor, we have reviewed the numerous works that explore the formation and destruction pathways of this molecular species in the ISM, from the point of view of astronomical observations, laboratory experiments, and theoretical calculations. While substantial progress on the subject has been achieved in the last few years, we are still far from having a full understanding on the physical conditions and chemical routes that lead to the formation/destruction of interstellar formamide. We summarise below some of the achievements, and accompany them by remaining questions and necessary steps to make progress.

\underline{Attempting a direct detection of formamide in a cold region:} It is clear that formamide is present in a few solar-mass protostellar cores, whether it is associated to the compact hot corino or to an outflow shocked region. How about the colder and less evolved pre-stellar cores? Or the cold external protostellar envelopes? So far, only upper limits or indirect estimates to the fractional abundance of formamide are available for these objects. Nonetheless, these detections would very strongly constrain the route formation of formamide, as the grain surface radical-radical formation would be extremely inefficient at such low temperatures because they would not be able to diffuse on the surface. Such a detection would therefore be the smoking gun of gas-phase synthesis of formamide\cite{balucani15}. While the expected amount of formamide in these cold objects is much lower than in hot corinos by 2 or 3 orders of magnitude, it would be very interesting to have a first direct detection, as is the case for other iCOMs\cite{bacmann12,vastel14,js16}. This requires dedicated long-integration single-dish observations with sensitive instrumentation such as the IRAM\,30-m telescope.

\underline{Towards improving the statistics:} While a relatively large number of high-mass molecular cores have been found to contain formamide, the sample is much more limited for solar-mass protostars. This calls for a statistical improvement. More protostellar sources should be targeted in order to evaluate how common it is to find formamide emission in these objects, as so far only protostars containing hot corinos display detectable amounts. This is actually only part of what constitutes a more general issue in protostellar astrochemistry, which is that of understanding why there is such a diversity in the chemical composition among protostars, especially when it comes to organic molecules\cite{sakai13}. The question can be more generally stated as follows: did the Solar System experience a hot corino phase during its birth? This is yet to be answered. In the same vein, only one protostellar shock has been explored in detail as far as formamide is concerned, from where it is clear that these environments are excellent laboratories to test theoretical predictions. Adding other chemically active protostellar shocks to the list will help solidify the conclusions drawn for L1157-B1.

\underline{Study of deuterated formamide:} As discussed in Sect.\,\ref{deut}, the ratio of the deuterated forms of formamide with respect to NH$_2$CHO and also with respect to the expected reactants producing formamide provide very strong constraints on its formation route. Unfortunately, these values have only been measured towards one source, IRAS 16293-2422. Evidently, in order to have a full overview of when gas-phase or grain-surface formation dominates more measurements are needed. This needs very high sensitivity observations, which can today be accomplished with last-generation facilities such as ALMA and NOEMA interferometers.

\underline{Considering other environments:} We have seen that formamide is not exclusively present in molecular cores where star formation is on-going. The detection of this species in diffuse molecular clouds\cite{corby15,eyres18}, where gas-phase chemistry dominates, and whose physical conditions differ from those found in SFRs, can place stringent constraints on the formation mechanisms of this molecule in such conditions, thus yielding a wider understanding of formamide chemistry in the ISM. Can the current known network of chemical reactions involving formamide explain the abundances measured in diffuse clouds? This remains to be answered.

\underline{Detecting formamide in interstellar ices}: While this has not been specifically addressed in the present review, tentative detections of formamide on the icy mantles of dust grains have been reported in two high-mass sources, NGC 7538 and W33A\cite{jones11,raunier04,schutte99}, at infrared wavelengths. Quantifying the amount of solid formamide present is difficult due to broad overlapping features with other species, such as ammonia (NH$_3$). Despite these uncertainties, it is worth trying to answer the question whether infrared observations with, for instance, the future James Webb Space Telescope (JWST), could deliver meaningful constraints on the amount of formamide on ices in different kinds of objects. This may require specific predictive calculations, both from spectroscopic and instrumental points of view.

\underline{Calculations and experiments, what are the next steps?} Answering the question whether gas-phase or grain-surface chemical reactions dominate the formation of formamide and in what conditions requires more experiments and theoretical calculations. On the former, even though very challenging, the measurement in the laboratory of the gas-phase reaction formaldehyde plus amidogen would be a crucial step, as theoretical calculations are unable to answer whether the barrier(s) of this reaction are submerged or not. On the grain-surface front, it would be important to carry out experiments on the deuteration of formamide: this is not particularly challenging and could provide strong constraints too. At the same time, laboratory experiments may elucidate the route proposed by Rimola and collaborators \cite{rimola18} concerning the formation via the reaction of frozen CN with the water molecules of the ice. Finally, nobody has so far addressed experimentally or theoretically the destruction routes of formamide: they are as important as the formation ones, if one wants to have the full picture.

\underline{Strength in numbers!} Last but not least, if there is one message to take home, it is that joining forces is the only way forward. In order to avoid mis-interpreting the observational data, it is imperative that communication flows between observers, modellers, experimental chemists and theoretical chemists, and that each of these groups keep feeding each other with new results and challenges for the future. Ending with a positive note, the authors of this review are all part of a large interdisciplinary project funded under the European Community: Astro-Chemical Origins (ACO\footnote{https://aco-itn.oapd.inaf.it/home}), the ideal framework to address not only the chemistry of interstellar formamide, but also many other fascinating astrochemical topics!

\begin{acknowledgement}

We warmly thank the reviewers, who provided pertinent corrections and suggestions to improve this review. This project has received funding under the European Union's Horizon 2020 research and innovation programme from the European Research Council (ERC) , for the Project ``The Dawn of Organic Chemistry" (DOC), grant agreement No 741002, and from the European MARIE SKLODOWSKA-CURIE ACTIONS  for the Project ``Astro-Chemistry Origins" (ACO), Grant No 811312. Claudio Codella thanks the PRIN-INAF 2016 ``The Cradle of Life -- GENESIS-SKA (General Conditions in Early Planetary Systems for the rise of life with SKA)", and  the program PRIN-MIUR 2015 STARS in the CAOS - Simulation Tools for Astrochemical Reactivity and Spectroscopy in the Cyberinfrastructure for Astrochemical Organic Species (2015F59J3R, MIUR Ministero dell'Istruzione, dell'Università della Ricerca e della Scuola Normale Superiore). This work was also supported by the Programme National ``Physique et Chimie du Milieu Interstellaire" (PCMI) of CNRS/INSU with INC/INP co-funded by CEA and CNES.

\end{acknowledgement}


\bibliography{lopez-sepulcre-nh2cho.bib}

\end{document}